\documentclass[%
 reprint,
 amsmath,amssymb,
 aps,prl
]{revtex4-2}

\usepackage{graphicx}
\usepackage{dcolumn}
\usepackage{bm}
\usepackage{hyperref}
\usepackage[mathlines]{lineno}

\begin{document}

\preprint{APS/123-QED}

\title{High efficiency uniform wakefield acceleration of a positron beam using stable asymmetric mode in a hollow channel plasma}

\author{Shiyu Zhou}
\affiliation{Department of Engineering Physics, Tsinghua University, Beijing 100084, China}
\author{Jianfei Hua}
\affiliation{Department of Engineering Physics, Tsinghua University, Beijing 100084, China}
\author{Wei Lu}
\email[]{weilu@tsinghua.edu.cn}
\affiliation{Department of Engineering Physics, Tsinghua University, Beijing 100084, China}
\author{Weiming An}
\affiliation{Beijing Normal University, Beijing 100875, China}
\author{Warren B. Mori}
\affiliation{University of Los Angeles, Los Angeles, California 90095, USA}
\author{Chan Joshi}
\affiliation{University of Los Angeles, Los Angeles, California 90095, USA}

\date{\today}

\begin{abstract}
  Plasma wakefield acceleration in the blowout regime is particularly promising for high-energy acceleration of electron beams because of its potential to simultaneously provide large acceleration gradients and high energy transfer efficiency while maintaining excellent beam quality. However, no equivalent regime for positron acceleration in plasma wakes has been discovered to-date. We show that after a short propagation distance, an asymmetric electron beam drives a stable wakefield in a hollow plasma channel that can be both accelerating and focusing for a positron beam. A high charge positron bunch placed at a suitable distance behind the drive bunch can beam-load or flatten the longitudinal wakefield and enhance the transverse focusing force, leading to high-efficiency and narrow energy spread acceleration of the positrons. Three-dimensional quasi-static particle-in-cell (PIC) simulations show that over 30\% energy extraction efficiency from the wake to the positrons and 1\% level energy spread can be simultaneously obtained, and further optimization is feasible.
\end{abstract}

\maketitle


High energy electron-positron ($e^-e^+$) colliders are highly desirable for precision studies of the Higgs Boson and discovering physics beyond the Standard Model \cite{Lou2019,Benedikt2019}. With the limitation on accelerating gradient imposed by breakdown at tens of MV/m, current radio-frequency accelerators are close to the maximum cost/size limit of such colliders operating at the energy frontier of particle physics. This has spurred intense research on novel acceleration schemes with much higher acceleration gradients and wall plug efficiency. One promising candidate is Plasma wakefield acceleration (PWFA) driven by a charge particle bunch \cite{Chen1985,Rosenzweig1987,Lu2006PRL}. It has not only demonstrated accelerating gradients of tens of GV/m, but also acceleration of a narrow energy spread $e^-$ bunch with high efficiency \cite{Litos2014,Joshi2020}. In PWFA, the relativistic drive bunch can be either electrons, positrons or protons \cite{Blumenfeld2007,Blue2003,Corde2015a,Doche2017,Adli2018}. Most groundbreaking results to-date have been obtained using an intense $e^-$  bunch to excite a nonlinear wake that accelerates a second $e^-$ bunch. However, this approach is not effective for $e^+$ acceleration because the volume at the very back of such wakes where the wakefield is both accelerating and focusing for $e^+$ is extremely small. Various methods have been proposed to overcome this limitation. Due to the uniform accelerating field in transverse planes and zero focusing force inside channel, wakes in hollow plasma channels produced by more easily available $e^-$  beams have been suggested for $e^+$ acceleration \cite{Chiou1995,Chiou1998,Schroeder1999,Schroeder2013,Gessner2016}. However, any misalignment of the drive and/or trailing bunches induces a strong beam-breakup (BBU) instability that leads to beam emittance growth and ultimately loss of positrons. This has limited further work on hollow channels for accelerating $e^+$ to high energy \cite{Lindstrom2018}. Recently, a hollow $e^-$ beam was proposed for $e^+$ acceleration in a uniform plasma, that creates a thin filament of plasma electrons on the axis that can focus and accelerate $e^+$ simultaneously. But its energy transfer efficiency appears to be very limited and it might be prone to kinetic instabilities \cite{Jain2015}. In another proposal, a wake excited by an $e^-$ beam in a finite-radius plasma column was proposed for $e^+$ transport and acceleration \cite{Diederichs2019}, where a narrow plasma electron filament is formed after the first wake cavity. However, the accelerating field for $e^+$ beam is highly nonlinear in transverse plane, limiting the final $e^+$ beam quality and the amount of accelerated charge seems limited. In this Letter, we propose a nonlinear scheme that provides field structures suitable for stably accelerating and focusing a positron bunch inside a hollow plasma channel. In this scheme the focusing field varies nearly linearly and the longitudinal field is almost independent of the transverse dimension throughout the channel in the region where the $e^+$ beam is placed. Such a field structure enables guided propagation and acceleration of a high charge $e^+$ beam in hollow plasma channel. Furthermore, the $e^+$ beam efficiently loads the wake and gains energy while maintaining a narrow energy spread. To our knowledge this is the first time a viable beam loading scenario for positrons in an electron beam driven plasma wake has been proposed that provides high beam quality and high efficiency.

\begin{figure}[ht]
  \includegraphics[width=8.6cm]{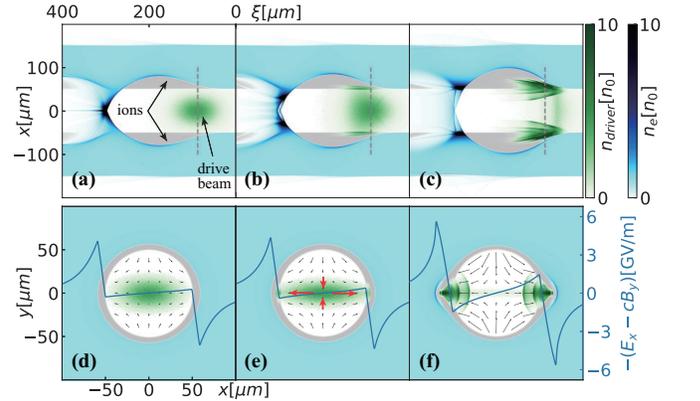}
  \caption{Evolution of an asymmetric electron beam and wakefields in a hollow plasma channel. (a-c) Plasma and beam densities in $x-\xi$ plane (Beam moves from left to right.) and (d-f) at the cross-section of beam center slice denoted by the gray dash line and transverse wakefields. Black arrows are the vector flow of transverse wakefield $-\vec{W}_\perp = -(E_x-cB_y)\hat{x} - (E_y+cB_x)\hat{y}$. Blue line is the lineout of $-\vec{W}_\perp$ at $y=0$. Red arrows indicate the spot size evolution of drive beam. The propagation distances are 0, 2.5, 10cm from left to right.}
  \label{Fig:1}
\end{figure}

We first summarize how our scheme works. The trick is to excite a quadrupole transverse wakefield in the hollow plasma waveguide using an asymmetric ($\sigma_x \neq \sigma_y$) electron beam driver. Here $\sigma_{x,y}$ are the beam RMS spot sizes in two transverse directions. As the beam propagates into the channel, its spot size evolution is determined by self-excited quadrupole wakefields as shown in FIG. \ref{Fig:1}. The narrow part of the beam is focused while the wider part is defocused until most electrons reach the inner plasma channel boundary in this plane. In the other plane the electrons are tightly focused. A dense drive bunch will repel the plasma electrons from the wall while leaving behind the much more massive ions. Once this state is reached, the electron beam propagates with little further evolution because the defocusing quadrupole fields are balanced by focusing forces from the exposed ions. These ions subsequently pull back the plasma electrons forming a sheath \cite{Sahai2016}. This process is analogous to formation of a half-bubble on opposite sides of the channel wall except that the returning sheath electrons overshoot the initial channel boundary and fill the entire cross-section of the hollow channel, providing focusing force that guides the trailing $e^+$ bunch in both planes if it is placed in this region. This $e^+$ bunch can extract a considerable amount of energy by flattening or beam loading the longitudinal accelerating field. Some plasma electrons are attracted towards the axis by the positrons and form a denser electron filament that enhances the focusing force. We will give a conceptual framework based on theory and then demonstrate it by 3D QuickPIC \cite{Huang2006,An2013} simulations.

Consider an ionized hollow plasma channel with density $n_p (r)=n_0 H(r-r_0)$, where $H$ is the step function, $r_0$ is the channel radius and $n_0$ is density of electrons. The wakefields driven by a relativistic point charge can be decomposed into discrete azimuthal modes \cite{Schroeder1999}. For a bunch of drive particles, we rearrange the transverse wake-functions by moments of beam position.

\begin{gather}
  \vec{W}_{\perp 0}(x,y,\xi) = 0
  \label{eq:transverse-wakefields-0-order}\\
  \vec{W}_{\perp 1}(x,y,\xi) = \lambda \hat{W}_{\perp 1}(\xi) [\langle x \rangle \hat{x} + \langle y \rangle \hat{y}]
  \label{eq:transverse-wakefields-1-order}\\
  \begin{aligned}
    \vec{W}_{\perp 2}(x,y,\xi) = \lambda &\hat{W}_{\perp 2}(\xi) [\langle x^2 - y^2 \rangle (x\hat{x} - y\hat{y}) \\ 
    & + \langle 2xy \rangle (y\hat{x} + x\hat{y})]
    \label{eq:transverse-wakefields-2-order}
  \end{aligned}
\end{gather}

The above expressions describe the linear transverse wake-functions in the zeroth, first and second order respectively, where $\lambda$ is the beam charge per unit length, angle bracket is average of particles at a given slice, $\hat{W}_{\perp m}(\xi)(m=0,1,2\dots)$ are coefficients determined by the parameters of hollow channel \cite{Schroeder1999} and $\xi \equiv ct - z$. The overall wake-function is the summation of all modes, but usually these low-order terms suffice. The electromagnetic field is the convolution of wake-function and charge distribution along $\xi$. For practical finite-thickness plasma channel, these equations are still useful approximations \cite{Spencer}.

Inside the hollow channel, transverse wakefield vanishes to the zeroth order, and other terms depend on the beam distribution. An on-axis axisymmetric bunch can propagate in the channel without any deflecting force. But if it is misaligned, the corresponding dipole wakefield as indicated by Eq. \ref{eq:transverse-wakefields-1-order} deflects the beam toward the boundary. Instead, if the beam is transversely asymmetric with little offset, quadrupole wakefield can be dominant and focus the beam in one plane and defocus it in the other until it reaches the plasma wall. In radio-frequency or dielectric accelerators, the high order wakefields have similar effects but must be suppressed or corrected \cite{BNSdamping,Oshea2020}, because once a particle hits the wall, it is lost. Since plasma is an ionized medium, the relativistic beam particles hitting the boundary repel the plasma electrons and expose the plasma ions. A stable equilibrium can then be found where the quadrupole defocusing forces are balanced by the Coulomb restoring force exerted by the ions which ceases the deflection.

The evolution of an on-axis elliptical electron beam propagating in a hollow plasma channel is presented in FIG. \ref{Fig:1}. The QuickPIC simulations used a $400\times 400\times 500\mu m$ ($x, y, \xi$) simulation domain with $512\times 512\times 1024$ cells. The plasma has density $n_0 = 3.11 \times 10^{16} cm^{-3}$, inner and outer radii $50 \mu m$, $150 \mu m$, with skin depth $30\mu m$. This beam has charge 2nC, energy 5.11GeV, tri-Gaussian profile with $\sigma_x = 20\mu m$, $\sigma_y=10\mu m$, $\sigma_z=30\mu m$ and normalized emittances $\epsilon_{nx}=20\mu m\cdot rad$, $\epsilon_{ny}=10\mu m\cdot rad$. So $\langle x \rangle =\langle y \rangle = \langle xy \rangle = 0$ and $\langle x^2 - y^2 \rangle = \sigma_x^2-\sigma_y^2$, and the leading term of transverse wake-functions is $\vec{W}_{\perp 2}=\lambda \hat{W}_{\perp 2} (\xi) (\sigma_x^2-\sigma_y^2)(x\hat{x} - y\hat{y})$, which defocuses it in $x$ direction and focuses in $y$ as plotted in FIG. \ref{Fig:1}(d). The beam gradually expands in $x$ and compresses in $y$ to increase its ellipticity as in FIG. \ref{Fig:1}(e). This drive beam is dense enough to evacuate the plasma electrons and expose the massive ions but it does not punch through the channel wall. Instead, the Coulomb field of the exposed ions pull it back. This total internal guiding of the relativistic beam has been measured experimentally \cite{Muggli2001a,Muggli2001b}. These electrons subsequently stay close to the plasma boundary while those inside the channel continue to move outward. Finally, after propagation of 10cm, most drive beam electrons are trapped near the plasma boundary, forming a quasi-steady state structure as shown in FIG. \ref{Fig:1}(c, f) and thereafter propagate stably. The drive beam eventually becomes equally divided into 2 parts, still possessing symmetries across the $x$ and $y$ axes.

The formation of the quasi-steady state structure is robust as it occurs for other beams with asymmetric profiles in $x$ and $y$. Larger degrees of asymmetry usually leads to stronger quadrupole wakefield and a shorter distance before steady state is reached. Particularly, if the initial beam profile is close to the steady state profile, such as two bunches separated transversely, a stable state is obtained immediately once the beams enter the hollow plasma channel.

\begin{figure}[ht]
  \includegraphics[width=8.6cm]{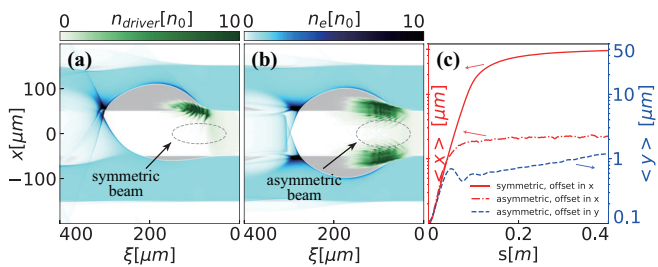}
  \caption{(a, b) Plasma and beam densities of a symmetric and asymmetric electron beam with initial offset $0.1\mu m$ in x direction after $20cm$ transportation in a hollow plasma channel. The dashed line is one-tenth-of-the-maximum contour of the initial beam density. (c) The evolution of the average offset for the electron beams in x and y directions.}
  \label{Fig:2}
\end{figure}

Transverse instability seeded by beam misalignment w.r.t. the channel axis is of major concern for hollow plasma channel. Here we show that the asymmetric (elliptic) beam is far more stable against the BBU instability than a round beam. FIG. \ref{Fig:2} illustrates the fully evolved (a) round and (b) elliptic beams and plasma profiles with initial centroid offset $0.1\mu m$ in $x$ direction. The round beam has $\sigma_r = 15\mu m$, $\epsilon_{nr}=15\mu m\cdot rad$ and other parameters the same as the above beam. The asymmetric beam is identical to that in FIG. \ref{Fig:1}. As aforementioned, a misaligned round beam excites transverse wakefield dominated by a dipole mode inside a hollow plasma channel, which kicks the beam towards the closest boundary. After propagation of 20cm this beam is totally deflected and hits the plasma wall. There is no place to load the witness $e^+$ bunch in this distorted plasma bucket. In contrast, the evolution of an asymmetric beam with some misalignment (FIG. \ref{Fig:2}(b)) is quite similar to the on-axis situation (FIG. \ref{Fig:1}(c)). At $z=20$cm, the electrons in the upper half-plane have charge 1.04nC, while initially that number is 1.004nC. The difference between the charge in two half-planes is slight compared to their total charge. FIG. \ref{Fig:2}(c) quantifies the evolvement of beam centroid in $x$ and $y$ directions. For the symmetric beam, centroid offset grows exponentially in initial offset direction until most electrons are stopped at the plasma boundary. The mean offset $\langle x \rangle$ for the elliptic beam grows at the beginning but saturates at a much smaller extent ($\sim 2\mu m$). When the asymmetric beam has $0.1\mu m$ offset in $y$ direction, the evolution is qualitatively the same, but the beam drifts slowly in this direction. After 40cm propagation, the centroid growth is about $1\mu m$. It may be possible to correct this offset using external magnets. Furthermore, asymmetric beam with other profiles can be explored to tolerate larger misalignments.

\begin{figure}[h]
  \includegraphics[width=8.6cm]{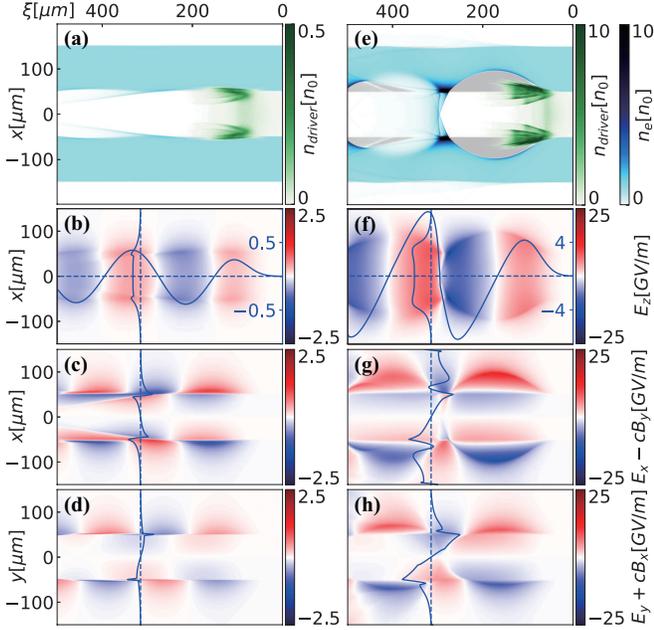}
  \caption{The unloaded wakefields. (a-d) is of a 100pC driver and (e-h) is of a 2nC driver. (a, e) The beam and plasma densities after stabilization. (b, f) $E_z$ field. Transverse wakefields in $x-\xi$ plane (c, g) and $y-\xi$ plane (d, h). The vertical lineouts of wakefields are at $\xi = 315\mu m$. }
  \label{Fig:3}
\end{figure}

However, stable propagation of the driver doesn’t ensure stable, efficient acceleration for $e^+$ beam. Guided $e^+$ transport and small emittance growth in plasma requires plasma electrons to flow through the positron bunch and provide a quasi-linear focusing force. In fact, when the driver (100pC $e^-$ bunch with other parameters same as in FIG. \ref{Fig:1}) is weak ($n_b<n_p$) it only slightly perturbs the plasma (FIG. \ref{Fig:3}(a)) and no plasma electrons are seen near the channel axis within the simulation box. The longitudinal field (FIG. \ref{Fig:3}(b)) is quasi sinusoidal on axis reaching a peak value of 400MV/m and nearly uniform in the transverse direction except near the slightly perturbed channel wall. The quadrupole wakefield dominates inside the channel (FIG. \ref{Fig:3}(c, d)), which changes sign in $y$ direction compared to the $x$ direction, so $e^+$ cannot be confined in both planes when it is loaded at the accelerating phase right behind the drive beam. The situation is dramatically changed when the driver strength is strong enough to excite a half-blowout-like wake on the channel wall on either side as shown in FIG. \ref{Fig:3}(e). In this 2nC driver case, the longitudinal wakefield is an order of magnitude larger, is non-sinusoidal, reaches a peak value of about 8GV/m and is nearly independent of the transverse position inside the channel (FIG. \ref{Fig:3}(f)). The returned plasma electrons which overshoot the channel boundary with some crossing the axis dominate the transverse wakefield structure inside the channel. As shown in FIG. \ref{Fig:3}(g, h), there is now an approximately linear focusing force for positrons in both dimensions, offering the possibility of positron acceleration with mild emittance growth.

Also note that the transverse wakefield is quadrupole in the first bucket where the drive beam resides as expected. However, in about half-wavelength-long region coinciding with where the plasma electrons are observed inside the channel in FIG. \ref{Fig:3}(e) in the second bucket, the transverse wakefield is focusing for $e^+$ and has a similar magnitude in both transverse planes (FIG. \ref{Fig:3}(g, h)). This is the region (around the dash line at $\xi=315 \mu m$ in FIG. \ref{Fig:3}(f-h)) where fields are both accelerating and focusing that is suitable for beam-loaded $e^+$ acceleration. 

Now we elaborate upon the physical effects seen in FIG. \ref{Fig:3}. For plasma electrons that are injected into the channel, the relationship between its transverse and longitudinal position is given by $\frac{dr}{d\xi} = \frac{v_r}{c-v_z}$, where $v_r$ and $v_z$ are the radial and longitudinal velocities respectively. The wavelength of the longitudinal wakefield is roughly $\lambda_0 \sim 2\pi / k_p$ \cite{Schroeder1999}. From FIG. \ref{Fig:3} we find that at the beginning of the positive $E_z$, those electrons are near the channel inner boundary. In order to load the positron beam at high gradient position, we desire $\left|\overline{\frac{dr}{d\xi}} \right| > \frac{r_0}{\lambda_0 / 4} = \frac{2k_p r_0}{\pi}$. For parameters of the above channel, it is around 1.05, which means the plasma electrons should be relativistic before entering into channel. This is normally satisfied when we are in the nonlinear regime of PWFA as in a uniform plasma with $\Lambda \sim 1$, where $\Lambda = \int_0^{\infty}k_p^2 r n_b/n_p dr$ is the normalized charge per unit length of the drive beam \cite{Lu2006POP,Lu2006PRL}. For the 2nC and 100pC drive beams, the peak $\Lambda$ is 0.9 and 0.045 respectively, consistent with the analysis.

We then employ 3D PIC simulations using QuickPIC to explore the positron beam loading scenario. High-efficiency positron acceleration inevitably involves strong interaction between positrons and plasma electrons, which will affect the wake to a great extent but in a complicated way. Surprisingly, we find that an intense $e^+$ beam does change the wakefield beneficially. Positrons attract plasma electrons inwards and form an electron filament on the axis to augment the focusing force. Besides, when the positron beam is loaded at a proper phase, the longitudinal wakefield can be flattened, and high-efficiency uniform acceleration is achieved.

\begin{figure}[h]
  \includegraphics[width=8.6cm]{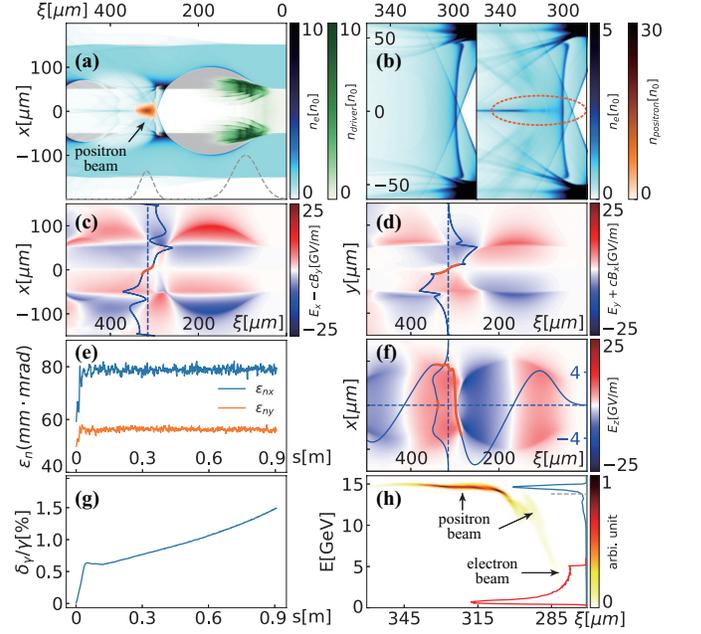}
  \caption{Positron beam loading effect and final beam quality. (a) Plasma and beam densities with positron beam loading, the dashed line is the current profile of two beams. (b) Plasma density in region of interest. Left: unloaded, right: beam loading case. The dashed line is one-tenth-of-the-maximum contour of the initial positron bunch density. (c, d) Loaded transverse wakefields. (e) Normalized slice emittance of the witness beam at $\xi = 315\mu m$ vs. propagation distances. (f) $E_z$ field. (g) Evolution of the energy spread of the witness beam with $\xi > 305\mu m$. (h) Final energy spectra of the two beams and longitudinal phase-space of the witness beam, the dashed line represents energy of 13.8GeV. The orange lines in (c, e, f) denote the wakefields within the range of $\pm 2\sigma_{x,z}$ for the positron beam.}
  \label{Fig:4}
\end{figure}

The positron beam loading effects once the 2nC, 5.11GeV elliptic driver bunch is stabilized are illustrated in FIG. \ref{Fig:4}. This $e^+$ beam contains 0.64nC charge, initial energy 10.2GeV, RMS length $15\mu m$, and $\sigma_x=5\mu m$, $\sigma_y=4\mu m$ and the emittances $\epsilon_{nx} = 60\mu m$, $\epsilon_{ny} = 50\mu m$, that is close to the $e^+$ beam used in SLAC's FACET experiment \cite{Corde2015a}. The two beams were injected from the outside with a fixed separation of $225\mu m$. As in FIG. 4(a, b), the majority of positrons are located at the focusing region and are confined, while a small fraction of beam head (front of the dashed red ellipse in FIG \ref{Fig:4}(b)) are in the first bucket and get deflected by the quadrupole transverse field. The deflection will cease when the beams arrive at the inner boundary and are refracted by the high density plasma electrons around the wall. So the beam charge is conserved during acceleration. An initially tailored current profile will help avoid this beam degradation. FIG. \ref{Fig:4}(b) compares the plasma electron distribution inside the hollow channel for the unloaded and loaded cases. For the unloaded situation these electrons nearly uniformly fill the cross-section of the channel, while the intense transverse electric field of the $e^+$ beam confines some electrons to form an on-axis electron filament. It is similar to the ion filament formation of blowout regime in uniform plasma with extremely intense electron beam \cite{Rosenzweig2005,An2017}. This filament has smaller radius than the initial positron beam size and will vary with the $e^+$ beam profile and charge. Since the plasma electrons are not uniformly distributed in the transverse plane, the focusing field is no longer linear within the positron bunch and becomes stronger near the axis (FIG. \ref{Fig:4}(c, d)). The evolution of the slice normalized emittance of the witness beam at beam center $\xi=315\mu m$ is illustrated in FIG. \ref{Fig:4}(e). The emittance grows by 20\%-30\% during the first 5cm propagation when the drive beam evolves, and stays constant for the rest. If the $e^+$ beam is injected after the driver is stabilized, the emittance growth should be strongly reduced. Previous study also finds that an overfocused beam can further reduce the emittance growth under this type of nonlinear focusing force \cite{An2017}. The filament remains after the positron beam passes, which extends the focusing region at $y$ direction.

Longitudinal wakefield shows the characteristic flattening due to beam loading by the positron beam. These loaded positrons shape the profile of the $E_z$ field by sucking in the returned plasma electrons and extract a substantial amount of energy from the wakefield. With proper beam profile and loading phase, it is possible for positron beam to get uniform acceleration. The $E_z$ field in FIG. \ref{Fig:4}(f) confirms the flattening effect of the accelerating field, with the peak value decreased from 8GV/m to around 5GV/m. At transverse direction, $E_z$ is smaller near the axis because of the electron-positron interaction, similar to the positron driven wakefield in a uniform plasma \cite{Lee2001}. FIG. \ref{Fig:4}(g) plots the energy spread of the $e^+$ beam with $\xi > 305\mu m$ (75\% of the total charge) during acceleration. The energy spread increases from 0 to 0.7\% after the first 5cm, then slowly grows to about 1.5\% during the next 85cm. Final energy spectra are presented in FIG. \ref{Fig:4}(h). Some $e^-$ almost deplete their energy, and most $e^+$ are accelerated to high energy while those in the front lose energy. The final longitudinal phase-space shows a sharp chirp on the beam head and uniform acceleration for the rest, consistent with the $E_z$ profile. For $e^+$ with $E>$13.8GeV, these part of the positrons contain 0.49nC, and the mean energy is 14.6GeV corresponding to an average gradient of 4.9GV/m, with an RMS energy spread of 1.6\%, and an energy transfer efficiency (the ratio of energy gained by this part to the energy transferred to the wake by the driver) of 33\%. The slice energy spread for uniform acceleration part is slightly smaller than the projected energy spread.

In summary, we have demonstrated a viable scheme of efficient high-gradient acceleration of a positron bunch with a narrow energy spread using an electron beam to drive the wake in a hollow plasma channel. This scheme appears promising and because of the nonlinear electron-positron interaction, there is a vast parameter space (beam profile, beam separation, transverse size, and emittance et al.) to explore and optimize the final beam quality.

This work is supported by the National Natural Science Foundation of China (NSFC) Grants (No. 11535006, No. 11991071, No. 11775125, and No. 11875175), CAS Center for Excellence in Particle Physics, and is also supported at UCLA by DE-SC0010064 and NSF grant 1734315.

\end{document}